\documentclass[preprint,pre,aps]{revtex4}
\input epsf
\usepackage{graphicx}
\usepackage{dcolumn}
\usepackage{bm}

\def\be{\begin{equation}}
\def\ee{\end{equation}}
\def\bea{\begin{eqnarray}}
\def\eea{\end{eqnarray}}

\def\bsea{\begin{subeqnarray}}
\def\esea{\end{subeqnarray}}

\begin{document}


\title{Nonequilibrium critical relaxations of the order parameter and 
 energy in the two-dimensional ferromagnetic Potts model}   

\author{Keekwon Nam${}^ 1$, Bongsoo Kim${}^ 1$ and Sung Jong Lee${}^ 2$}
\affiliation{${}^ 1$ Department of Physics, Changwon National University, 
Changwon 641-773, Korea\\
${}^ 2$ Department of Physics, The University of Suwon, Hwaseong-Si 445-743, Korea}


\date{\today}

\begin{abstract}

The static and dynamic critical properties of the ferromagnetic $q$-state 
Potts models on a square lattice with $q=2$ and $3$ are numerically studied
via the nonequilibrium relaxation method. The relaxation behavior of both the 
order parameter and energy as well as that of the second moments are investigated, 
from which static and dynamic critical exponents can be obtained.
We find that the static exponents thus obtained from the relaxation of the order 
parameter and energy together with the second moments of the order parameter exhibit 
a close agreement with the exact exponents, especially for the case of $q=2$ (Ising)
model, when care is taken in the choice of the initial states for the 
relaxation of the second moments. As for the case of $q=3$, the estimates for 
the static exponents become less accurate but still exhibit reasonable agreement 
with the exactly known static exponents. 
The dynamic critical exponent for $q=2$ (Ising) model is estimated from the 
relaxation of the second moments of the order parameter with {\em mixed} initial 
conditions to give $z(q=2) \simeq 2.1668(19) $.

\end{abstract}

\maketitle

\section{Introduction}

The so-called nonequilibrium relaxation (NER) method \cite{nito1,ozeki_nito1} deals with
the nonequilibrium critical relaxation of a statistical model instantaneously 
brought to its critical temperature from its nonequilibrium initial state, 
typically a fully ordered state. A unique feature about the NER method is that it 
allows one to determine the equilibrium critical properties of statisitical systems 
from their {\em nonequilibrium} relaxation kinetics. Therefore, together with the
 method \cite{bz1,lrz} using a short-time critical dynamic properties 
\cite{janssen,huse} (see Section III.D), the NER method has been an interesting and 
valuable tool to study the statistical systems whose critical properties are unknown. 
   
With recent advances in computing power, it seems worthwhile to reinvestigate 
in more depth the potentials or limitations of the NER method in terms of numerical 
accuracies in static and dynamic exponents. In this work we chose the ferromagnetic 
$q$-state Potts model \cite{fwu} on a square lattice (with $q=2$ and 
$3$) as a testground for the NER method. It appears to us that in applying the NER 
method, the energy relaxation itself has been overlooked in evaluating the critical 
exponents: previous works \cite{nito2,ozekito} employed mainly the second moments 
involving the energy, ignoring the relaxation of the energy itself. In this work we 
utilized the critical energy relaxation as well as that of the order parameter for 
evaluating the critical exponents. Due to better self-averaging of these single moment
quantities, higher accuracy is expected for the estimates of the exponents. Since for 
the ferromagnetic $q$-state Potts model the critical temperature and energy are exactly 
known for general values of $q$, and the static critical exponents are exactly known for 
$q \le 4$ \cite{fwu,baxter}, our work can be considered as a calibrational study on the 
NER method as to how accurate estimates for the critical exponents can be given via the 
NER method.

We find that by combining the relaxation of the two first-moment quantities, i.e., order 
parameter and energy one can obtain the ratio of the equilibrium critical exponents,
through which the nature of the phase transition or, equivalently, the universality class, 
may be determined regardless of our knowledge of the exact value of the dynamic exponent $z$. 
By incorporating the time dependence of the second moment of the order parameter (starting 
from either disordered or ordered initial states), the dynamic exponent 
can be independently obtained. This can be combined with the relaxation of the first moment 
quantities to give the two independent static exponents. We here find that the second 
moment of the order parameter obtained with disordered initial state gives the static 
exponents with higher accuracy. 
Alternative way of obtaining static exponents is to employ the second moments involving 
energy fluctuations. However, the energy fluctuation appears to be influenced by strong 
logarithmic corrections in the case of $q=2$, which makes it difficult to estimate the 
exponents with high accuracy. 

The present work is motivated by our interest in applying the NER method to the statistical 
systems whose critical properties are not completely understood. We are particularly 
interested in clarifying the long-standing controversy on the critical properties of the 
FFXY model \cite{ffxy} on a square lattice \cite{hasenbush} and the antiferromagnetic XY
 model \cite{kawamura} (or the antiferromagnetic clock model \cite{noh,surungan}) on a 
triangular lattice. Even though there exist previous studies \cite{hluo,qchen,mbluo,ozekito} 
from nonequilibrium dynamics perspective on this problem, we hope to undertake more 
careful studies on the FFXY models by applying the NER method with the strategy presented 
in this work.

\section{Model and Simulation method}

In the ferromagnetic $q$-state Potts model on a square lattice each spin $\sigma_i$ can 
take one of the $q$ possible values, $\sigma_i=1, 2, \cdots, q$, and the spin interaction 
is defined by the following Hamiltonian
\be
H=-J \sum_{<ij>} \delta(\sigma_i,  \sigma_j),  \qquad  \sigma_i=1, 2, \cdots, q
\ee
where $\delta(a,b)$ is the Kronecker delta function and $<ij>$ denotes the nearest neighbor 
spin pair. $J(>0)$ is the interaction strength. 

The model is known to exhibit a continuous phase transition for $q \le 4$ and discontinuous 
ones for $ q > 4$ in two dimensions \cite{fwu}. For $q=2$, the Potts model is equivalent 
with the Ising model since $\delta(\sigma_i, \sigma_j)=(1+S_i S_j)/2$ with $S_i=\pm 1$.
It is remarkable that though the model is not exactly solvable except for $q=2$, the critical
temperature $T_c$ and energy $E_c$ are exactly known for general $q$ in some two dimensional 
lattices. For the square lattice they are given by \cite{fwu}
\be
k_B T_c = \frac{J}{\ln(1+\sqrt{q})},  \qquad  
E_c=-J\left( 1+\frac{1}{\sqrt{q}} \right)
\ee
where $k_B$ is the Boltzmann constant. Here and after we set $k_B=1$ and $J=1$.  Hence 
$T_c=1.13459 \cdots$, $E_c=-1.70711 \cdots $ for $q=2$, and 
$T_c=0.99497 \cdots $, $E_c=-1.57735 \cdots $ for $q=3$. The exact static critical exponents 
are given in TABLE~I.

We will be interested in characterizing the time evolution of the system instantaneously 
'heated' to $T_c$ from a fully ordered initial state. The system then proceeds toward the 
equilibrium via Monte Carlo kinetics with the Metropolis algorithm for the flip of a 
randomly selected spin. Since we are interested in the long time dynamics of the system in 
the thermodynamic limit, we take the system size large enough so that the simulation 
can appropriately mimic the nonequilibrium relaxation of the infinitely large system within 
the simulation time window. This is similar to the case of phase-ordering kinetics \cite{abray} 
in which the equilibration time is practically infinite due to large system size, making the 
time evolution ever nonequilibrium. We use the square lattice with linear size $L=600$.
The maximum simulation time is $t_{max}=10^4$ Monte Carlo steps (mcs). The presented results 
are averages over $20,000$ samples. Compared to the existing dynamic studies 
\cite{bz1,schzh1,schzh2,lischzh,okano,zhang,ying} on the present model, our work has employed 
bigger lattice size with one or two decades longer Monte Carlo steps of simulations. Within 
the simulation time window, we have checked that there is no finite size effect on our simulation
results.  

\section{Measurements and discussions}

 We first define the local order parameter $m_i(t)$ and local energy $e_i(t)$ for the model, 
and their respective sums ${\cal M}(t)$ and ${\cal E}(t)$ as 
\bea
m_i(t) &\equiv& \frac{1}{(q-1)} 
\Big( q \delta(\sigma_i(t), \sigma_i(0))-1   \Big), \qquad
e_i(t) \equiv -\frac{1}{2} \sum_j^{(i)}  \delta(\sigma_i(t), \sigma_j(t)) \nonumber \\
{\cal M}(t) &\equiv& \sum_i m_i(t), \qquad {\cal E}(t) \equiv  \sum_i e_i(t)
\eea
where $\sum_j^{(i)}$ denotes the sum over the nearest sites of the site $i$. We have set 
$\sigma_i(0)=1$ in most of the present simulations. In the NER method, the determination of 
the critical exponents involves the cumulants of ${\cal M}(t)$ and ${\cal E}(t)$. 
The order parameter $M(t)$ and the energy density $E(t)$ are given by 
$M(t) \equiv  \big< {\cal M}(t) \big>/N$ and $ E(t) \equiv \big< {\cal E}(t) \big>/N$ 
respectively where $N$ is the total number of spins, and the angular bracket $<\cdots>$ 
denotes average over different realizations of the time evolution.

\subsection{Locating the critical temperature and energy}

The NER method provides us with an efficient tool to determine $T_c$ with high accuracy.
If the temperature $T$ is above or below $T_c$, then the order parameter would relax 
{\em (stretched) exponentially} in time to its equilibrium value which is zero for 
$T \ge T_c$, or  nonzero for $T<T_c$ in the continuous transition as in the present model.
The relaxation time scale $\tau$ depends on how close $T$ is to the critical temperature 
$T_c$. Approaching $T_c$, $\tau$ exhibits a power law divergence. Below $T_c$, the order 
parameter $M(t)$ relaxes toward the nonvanishing equilibrium value. Since only at $T_c$  
the order-parameter exhibits a power-law relaxation, $T_c$ can be located as the temperature 
at which the order-parameter relaxation gives the best straight line in a log-log plot of 
$M(t)$ versus $t$. More accurate location of $T_c$ naturally requires longer simulation 
time. 

Since  $T_c$ is exactly known for the present model, we can provide a test example how 
the order-parameter relaxation can be used to narrow down the critical temperature, as 
shown in Fig.~1(a) ($q=2$) and Fig.~1(b) ($q=3$). Figure~1a shows the relaxation of the 
Ising ($q=2$) order parameter for various temperatures near $T_c$. $M(t)$ shows  strong 
upward and downward curvatures at $T=1.13$ and at $T=1.14$, respectively. So the critical
 temperature should be in the range $1.13 < T_c <1.14$. We find in this way that the 
temperature range can be readily narrowed down to $\delta T=4 \times 10^{-4}$ within the 
present simulation time ($t_{max}=10^4$  mcs). That is, we have $1.1342 < T_c < 1.1350$.  
Higher accuracy  would  be achieved with longer simulation time. In the same manner, 
for $q=3$, $T_c$ can be easily located within the range $\delta T=5 \times 10^{-4}$ as 
demonstrated in Fig.~1(b).

In contrast to the case of the present system, the exact value of the critical energy 
is not known for many other systems. For such cases, using an ansatz of the critical energy
relaxation at $T_c$, one can determine $E_c$ by tuning the value of $E^{\ast}$ such that
 $\big(E^{\ast}-E(t)\big)$ versus $t$ gives the best straight line at long times in the 
log-log plot. Figure~2 shows such an example for the present system. In our simulation 
time window, one could readily narrow down $E_c$ within the range $\delta E \simeq 10^{-3}$. 

\subsection{Relaxations of the first moments}

We first measure the relaxation of the order parameter $M(t)$ and the energy density
$E(t)$. When the system is instantaneously  brought to $T_c$ starting from a fully
ordered state, both quantities exhibit power-law relaxations toward their equilibrium 
values \cite{siegert,stauffer0,nito2}. For better analysis of the relaxation behavior and 
estimation of the critical exponents, we included correction terms to the leading-order 
scaling behavior in the following form (see APPENDIX for a derivation of the leading-order 
critical relaxation)
\be
M(t) \sim t^{-\beta/z\nu} \Big(1+\frac{c_{M}}{t^{\delta_{M}}}+ 
\frac{c'_{M}} {t^{\delta'_{M}}}+ \frac{c''_{M}} {t^{\delta''_{M}}}+ \cdots \Big),  \qquad
E_c-E(t) \sim t^{-(\nu d -1)/z\nu}
\Big(1+\frac{c_{E}}{t^{\delta_{E}}}+ \frac{c'_{E}}{t^{\delta'_{E}}}+ 
\frac{c''_{E}}{t^{\delta''_{E}}}+ \cdots \Big)
\ee
where $z$ is the dynamic exponent, $\beta$ and $\nu$ the static exponents, and $d$  
the spatial dimension ($d=2$ in the present case). In (4), $c_{M}$, $c'_{M} $, $c''_{M}$ and 
$c_{E}$, $c'_{E}$, $c''_{E}$ are constants, and $\delta_{M}$, $\delta_{E}$, etc are the 
exponents of the correction-to-scaling terms. Simulation results for the relaxations of 
$M(t)$ and $(E_c-E(t))$ are shown respectively in Fig.~3(a) and Fig.~3(b). 

It is easy to see that the general form of Eq.~(4) with all the correction exponents kept 
independent, is not appropriate for fitting. For example, if we put all the correction exponents 
equal to one another, then we can generate the same function with many different combinations of
the coefficients provided that the sum of the coefficients are the same.      
This will cause the fitting procedure with a general initial guess to fail to converge to unique 
stable values for the values of exponents and other coefficients. 
Therefore, we should impose some constraints on the correction exponents to achieve a stable fit. 
One of the schemes we tried is to let the first correction exponent to be a free fitting parameter
but the $n$-th correction exponent are constrained to be equal to the first correction exponent 
plus $(n-1)$. That is, we put $\delta'_{M} = \delta_{M}+1 $ and $\delta''_{M} = \delta_{M}+2$, etc 
(which we call FIT$n$A, where $n$ refers to the number of the correction terms). 
We also tried a scheme where the first correction exponent to be a free fitting parameter but 
the $n$-th correction exponents are constrained to be equal to $n$ times the first correction 
exponent. That is, we put $\delta'_{M} =2 \delta_{M}$ and $\delta''_{M} = 3 \delta_{M}$, etc 
(which we call FIT$n$C). A special scheme of correction is to assume an analytic form for the 
correction terms where the correction exponents are all constrained to integer values (which we 
call FIT$n$B). The cases of $\delta_{M} =1 $ in FIT$n$A and FIT$n$C reduce to the scheme of
FIT$n$B. The special case of fitting with a simple power law without correction terms will be 
referred to as FIT$0$.

First consider the case of $q=2$. In this case, we incorporated up to third correction terms 
(i.e., FIT3A, FIT3C and FIT3B) in our fit. In the cases of FIT3A and FIT3C, for reasonable
convergence of the fitting procedure and estimation of the errorbars, we tuned the value of 
the coefficients of the highest order correction terms (i.e., $c''_M $ and $c''_E$). 
These values of  $c''_M $ and $c''_E$ 
were tuned (chosen) such that the resulting fit function shows good agreement with the data. 
For some finite range of these coefficient values ($c''_M$ ranging from $0.0$ to $0.1$), 
the quality of the fit is found to be acceptible. 
The median of the values of the exponents was chosen as the representative value of the exponent 
and the range of the values can be used to estimate the error bars. As for the relaxation of 
the magnetization, we found that, as the value of $c''_M $ was varied, the value of the dominant 
relaxation exponent $\beta /z\nu $ showed only small variation. We also found that, in terms 
of the value of the dominant exponent $\beta /z\nu $ and the first correction exponents, the two 
schemes of FIT3A and FIT3C gave almost the same results (see TABLE II). If we take the median 
of the fitted values from this analysis as the most probable value of $\beta /z \nu$, we get 
\be
\frac{\beta}{z\nu} = 0.057722(42) \qquad  \qquad \quad \mbox{for} \quad q=2. \nonumber 
\ee
Now the corresponding value of the first correction exponent $\delta_{M}$ from the above fit
exhibits a monotonic increase from $0.91$ to $1.06$ (FIT3C) and also from $0.90$ to $1.07$ (FIT3A) 
as the value of $c''_M$ increases from $0.0$ to $0.1$ (TABLE II). Taking the median of the 
combined range (from $0.90$ to $1.07$) we may take the representive value of $\delta_{M}$ as 
$\delta_{M} \simeq 0.985(85)$. This may indicate that the exact value of the first correction 
exponent is equal to one. Incidentally we also tried to fit the data with analytic corrections 
in which the first correction exponent is set to be equal to one, the second correction exponent 
two, etc. We found that the fitting quality was excellent with essentially the same dominant 
exponent as the representative value given above.

As for the relaxation of the excess energy, we also applied the above fitting schemes. 
However, in this case of excess energy, we had to directly tune the value of the first
correction exponent in order to obtain a reasonable fit to the data. We found 
that the best fit could be obtained in the range of the value of $\delta_E$ between 
$0.90$ and $1.05$ (see TABLE III) which again is in reasonable agreement with the case 
of the magnetization relaxation (Fig.~3(b)). The values of the fitted relaxation exponents
are 
\be
\frac{\nu d -1 }{z\nu} = 0.46227(71) \qquad  \qquad \mbox{for} \quad q=2. \nonumber
\ee

Now we turn to the relaxation dynamics of $q=3$ model. As for the magnetization relaxation, 
we tuned the value of the coefficient $c''_M $ or $c''_E $ of the highest order correction 
terms. We find that both schemes FIT3A and FIT3C give approximately the same results for 
values of the relaxation exponent $\beta /z\nu $ (TABLE IV). Combining the results of FIT3A 
and FIT3C we obtain the values of the exponent as 
\be
\frac{\beta}{z\nu} = 0.06020(17) \qquad  \qquad \quad \mbox{for} \quad q=3. \nonumber 
\ee
Correspondingly the value of the first correction exponent $\delta_{M}$ ranges between 
$0.65$ and $0.98$. 
As for the relaxation of the excess energy, we also applied the above fitting schemes. 
We found that the best fit could be obtained in the range of the value of $\delta_E$ 
between $0.6$ and $0.74$ which exhibit a narrower range for the correction exponent 
compared with the case of magnetization relaxation. The value of the fitted relaxation 
exponent for excess energy is (TABLE V)
\be
\frac{\nu d -1 }{z\nu} = 0.3626(40) \qquad  \qquad \mbox{for} \quad q=3. \nonumber 
\ee
Figures 3(c) and 3(d) show the time dependence of $M_A(t)\equiv M(t) A_0 t^{\beta/ z\nu}$
for $q=2$ and $q=3$ respectively together with the fitting functions.  
Here $A^{-1}_0 t^{-\beta/z \nu}$ represents the asymptotic scaling obtained from the 
fitting.  Also, Figs.~3(e) and 3(f) show the time dependence of 
$E_A(t) \equiv (E_c-E(t)) B_0 t^{(\nu d -1)/z \nu}$ with the fitted values of the 
exponents for $q=2$ and $q=3$ respectively together with the fitting functions, 
demonstrating the asymptotic nature of the scaling behaviors. 
Here $B^{-1}_0 t^{-(\nu/d -1)/z \nu}$ represents the asymptotic scaling obtained from the 
fitting.  

Note that the value of $\beta/z\nu$ is particularly small. This implies that the 
order-parameter relaxes  much more slowly than the energy does: even after four 
decades of relaxation time, the order parameter relaxed only half of its initial 
value. This is another reason why the statistics is better for the order parameter 
relaxation than for the energy relaxation. The above values of $\beta/z\nu$ for 
both $q=2$ and $3$ can be compared with the results obtained from the short-time 
dynamics both on square \cite{bz1,schzh2} and triangular \cite{ying} lattices.

Since $\beta/z\nu$ and $(\nu d-1)/z\nu$ contain the common factor $1/z\nu$ which involves 
the dynamic critical exponent $z$, taking the ratio of the former to the latter eliminates
the exponent $z$, yielding $\beta/(\nu d-1)$ which involves only the two static exponents.
Then the numerical values of the ratio $\beta/(\nu d-1)$ obtained from the above fit 
can be compared with the corresponding exact values as 
\bea
\frac{\beta}{\nu d-1}&= & 0.1249(4), \qquad \quad \, \,
\Big(\frac{\beta}{\nu d-1}\Big)_{\mbox{exact}}=\frac{1}{8}=0.125 
\quad \mbox{for} \quad q=2,  \nonumber \\
\frac{\beta}{\nu d-1}&= & 0.1660(23), \qquad  \quad
\Big(\frac{\beta}{\nu d-1}\Big)_{\mbox{exact}}=\frac{1}{6}=0.16667
 \quad \mbox{for} \quad q=3.
\eea
We see that for both cases of $q=2$ and $q=3$, the values of ${\beta}/(\nu d -1)$
are quite close to the exact theoretical values.

Elimination of the dynamic exponent $z$ in the above ratio is an important feature 
that may be used to distinguish the universality class of systems whose critical
properties are not known. Prominent examples of such systems are the two-dimensional 
FFXY models \cite{hasenbush}. We point out that as is done here, measuring the 
relaxations of the chirality order parameter and the energy for such systems 
one can determine whether the chirality transition in that system belongs to the Ising 
or the $3$-state Potts universality class, or to another universality class, 
without knowing all the static exponents separately. 


One can obtain an estimate for $z$ from the fitted exponents for the relaxation of
magnetization and energy as given above by making use of the fact that the static 
exponents are exactly known in the case of $q=2$ and $q=3$. This is perhaps one of 
the easiest ways of obtaining $z$ for systems whose static critical properties are exactly 
known \cite{nito3,murase_nito}. If we use the magnetization relaxation, the values for 
$z$ derived from the values of the relaxation exponents in (5)-(8) (assuming the exact 
theoretical values of the static exponents) are given by 
\bea
z_M &=& 2.1656(16) \qquad \qquad   \, \,  
z_E = 2.1632(33) \quad \mbox{for} \quad q=2, \nonumber \\
z_M &=& 2.2150(62) \qquad  \qquad  z_E =2.2064(110) \quad \mbox{for} \quad q=3,
\eea
Here values of the exponents are the median values in the optimal fitting regime, 
while the error bars are estimated from the maximal dispersion of the fitted values.   


\subsection{Relaxations of the second moments}

The above procedures of obtaining $z$ of course cannot be carried out for systems 
with unknown critical properties. It is thus desirable to provide ways of 
independently measuring the dynamic exponent. We present below one systematic 
way of doing it by considering the second moments of the order parameter, 
$C_{{\cal M}{\cal M}}(t)$, which exhibits a leading-order scaling behavior 
in time \cite{nito2} as (see APPENDIX)
\be
C_{{\cal M}{\cal M}}(t) \equiv N 
\Big[ \big< {\cal M}^2(t)\big> -\big<{\cal M}(t) \big>^2 \Big] 
 \sim t^{(d-2\beta/\nu)/z}
\ee 
Note that (11) is valid for {\em disordered} initial states as well. The dynamic 
exponent $z$ can be isolated if one considers a time-dependent second moment 
$f_{{\cal M}{\cal M}}(t)$ defined as
\be
f_{{\cal M}{\cal M}}(t) \equiv \frac{C_{{\cal M}{\cal M}}(t)}{M^2(t)}
 \sim t^{d/z} 
\ee
which is obtained using the leading-order behavior for $M(t)$ given in (4).
One usually measures $f_{{\cal M}{\cal M}}(t)$ in two ways: one measures 
$f_{{\cal M}{\cal M}}(t)$ for fully ordered initial states, or one can first 
measure $C_{{\cal M}{\cal M}}(t)$ for disordered initial states, and then 
divide it by $M^2(t)$ measured for the fully ordered initial states. 
The latter way is therefore using the two different types of initial conditions. 
We denote the former result by $f_{{\cal M}{\cal M},ordered}(t)$, and the 
latter one by $f_{{\cal M}{\cal M},mixed}(t)$. Alternative way \cite{bz1} 
of obtaining $z$  using the short-time critical dynamics is described in 
the next section.

Shown in Fig.~4(a) and Fig.~4(b) are $f_{{\cal M}{\cal M}}(t)$ for $q=2$ and 
$q=3$, respectively. Here, we apply similar methods of analysis as in the 
previous section. 
That is, $f_{{\cal M}{\cal M}}(t) \sim t^{d/z} 
\left(1+c_{{\cal M}{\cal M}}/t^{\delta_{{\cal M}{\cal M}}} + .....\right)$, 
We found that for the case $q=2$ (using both FIT2A and FIT2C)
$f_{{\cal M}{\cal M},ordered}(t)$ and $f_{{\cal M}{\cal M},mixed}(t)$ give 
(TABLE VI and VII)
\be
z_{ordered} =2.1545(52), \qquad z_{mixed}=2.1668(19)  \quad \mbox{for} \quad q=2. 
\ee
We see that there exists small but non-negligible discrepancy between the exponents 
$z_{mixed}$ and $z_{ordered}$. We also note in particular that $z_{mixed}$ is very
close to $z_M$ obtained from the relaxation of the magnetization in (5) (assuming the 
exact values of the static exponents). 
Figure~4(c) shows the time dependence of the
quantity $f_{{\cal M}{\cal M},A}(t)\equiv f_{{\cal M}{\cal M}}(t) C_0 t^{-d/z}$ 
for the case of mixed initial states (for $q=2$) together with the fitting 
function, where $C^{-1}_0 t^{d /z}$ represents the asymptotic scaling obtained 
from one of the best fits.  

In the case of $q=3$, $z_{ordered}$ and $z_{mixed}$ exhibit a larger discrepancy:
\be
z_{ordered} = 2.1334(35),  \qquad z_{mixed}=2.1735(40) \quad \mbox{for} \quad q=3
\ee
As first pointed out by Zheng \cite{bz1}, this difference between the two estimates 
is not due to the statistical error of the data. This difference is also observed in 
triangular lattice \cite{ying}. More severe disagreement between $z_{ordered}$ and 
$z_{mixed}$ was reported for the Baxter-Wu model 
\cite{wubax,adler}: $z_{ordered}=2.07(1)$ \cite{santos} and 
$z_{mixed}=2.294(6)$ \cite{arashiro}. We suspect that the underlying reason for this 
discrepancy is that the time scale at which the scaling behavior sets in, may be much 
longer for $f_{{\cal M}{\cal M},ordered}(t)$ than that for $f_{{\cal M}{\cal M},mixed}(t)$ 
as $q$ gets larger, which may be due to the nature of the broken symmetry of the initial
states related to the higher degeneracy of the ground states for larger $q$. It is thus 
expected that the  stronger discrepancy will be observed for the $4$-state Potts model 
as well. 

\subsubsection{The static exponents $\nu$ and $\beta$}

The values of dynamic critical exponents obtained from second moments of the
order parameter can now be combined with the relations of exponents obtained
from the relaxation of the first moments of the order parameter and energy to
give the static exponents such as $\nu$ and $\beta$. 
Since it appears that the value of $z_{mixed}$ (especially for $q=2$) is more 
consistent with the results of magnetization and energy relaxation, we 
substituted $z_{mixed}$ for the dynamic exponents in the relaxation of 
magnetization and energy. 
For example in the case of $q=2$, substituting the value of $z_{mixed}=2.1668(19)$ 
in (13) for $z$ in the relations 
(5) and (6) gives $1/\nu=0.9984(25)$ and $\beta=0.12527(51)$. These values yield
 close estimates to the exact exponents: $1/\nu=1$ and $\beta=1/8=0.125$.  


As for $q=3$, when the value of $z_{mixed}=2.1735(40)$ in (14) is substituted 
for $z$ in the relations (7) and (8) 
we obtain $1/\nu=1.212(16)$ and $\beta=0.1080(20)$, which appears to exhibit 
approximate agreement with the exact values $1/\nu=6/5=1.2$ and $\beta=1/9=0.1111$. 
But this is definitely less accurate than the case of $q=2$. If we use $z_{ordered}$ 
instead of $z_{mixed}$, then we obtain even less accurate results for the 
static exponents, which is naturally expected from the discrepancy of the value
of $z_{ordered}$ from $z$ obtained from the relaxation of order parameter or energy
assuming the exact static exponents. One thus should be cautious in determining 
$z$ from the second and higher moments of the order parameter for systems with 
unknown critical properties. 

We have so far followed the procedure of using the relaxations of the order parameter, 
the energy, and the second moment of the order-parameter. An alternative way 
\cite{nito2,ozekito} of obtaining the exponent $\nu$ is to use the second-order moments 
involving energy fluctuations, $f_{{\cal M}{\cal E}}(t)$ or $f_{{\cal E}{\cal E}}(t)$, 
 which exhibit the leading-order scaling behaviors as \cite{nito2} (see APPENDIX)
\bea
f_{{\cal M}{\cal E}}(t) &\equiv& N \left[ \frac{<{\cal M}(t) {\cal E}(t)>
-<{\cal M}(t)> <{\cal E}(t)>} 
{<{\cal M}(t)><{\cal E}(t)>} \right] \sim t^{1/z\nu}, \nonumber \\
f_{{\cal E}{\cal E}}(t) &\equiv& N \left[ \frac{<{\cal E}^2(t)>-<{\cal E}(t)>^2}
{<{\cal E}(t)>^2} \right] \sim t^{\alpha /z\nu}
\eea
For $q=2$, we first obtain the leading-order exponent $1/z\nu=0.4624(29)$ by using 
FIT3A (Fig.~5(a)). However, we are unable to extract the exponent $(2-\nu d)/z\nu$ 
from $f_{{\cal E}{\cal E}}(t)$ since  it exhibits a strong curvature with the local slope
decreasing in time, as shown in Fig.~5(b). This may indicate that the exponent $(2-d\nu)$ 
eventually vanishes in the long time limit, implying $\nu=1$. We suspect that this 
strong curvature in the log-log plot of $f_{{\cal E}{\cal E}}(t)$ may correspond 
to a logarithmic correction. Figure~5(c) shows such a fit with 
$f_{{\cal E}{\cal E}}(t) \sim (\log t)^{\phi}$ with $\phi \simeq 0.80$. 
From the value of the above exponent $1/z\nu=0.4624(29)$ 
we obtain $1/\nu=1.0019(72)$ using $z_{mixed}=2.1668(19)$. This estimate compares quite 
well with that obtained from $M(t)$, $E(t)$, and $f_{{\cal M}{\cal M}}(t)$.

For $q=3$, in contrast to the Ising ($q=2$) case, both moments $f_{{\cal M}{\cal E}}(t)$ 
and $f_{{\cal E}{\cal E}}(t)$ exhibit asymptotic power-law behavior. 
The average slopes obtained from Fig.~5(a) and Fig.~5(b) are given by 
$1/z\nu=0.5515(61)$ and $(2-\nu d)/z\nu=0.1911(72)$. These values yield the static and 
dynamic exponents $1/\nu=1.210(58)$ and $z=2.193(129)$. It is interesting to see that
this dynamic exponent is larger than $z_{mixed}=2.173(2)$ given in (14), and is rather 
closer to those given in (10). However, one should also note that the error bars are 
much larger in this case.  


One can therefore conclude that for the present system using the relaxation of the order
 parameter and the energy combined with the value of $z_{mixed}$ yields better estimates 
for static exponents than using the second-order moments involving the energy fluctuations. 
From this point of view, the moments $f_{{\cal M}{\cal E}}(t)$ and $f_{{\cal E}{\cal E}}(t)$ 
may be used to check the consistency of the results obtained using the energy relaxation 
itself.

\subsection{The short-time critical dynamics}

At this point, for comparison, it is worthwhile to briefly describe the method using the 
short-time critical  dynamics  which offers an alternative way of determining the static 
and dynamic exponents. More details can be found in \cite{bz1}.  When quenched to $T_c$ 
from the initial disordered state but with sufficiently small 'magnetization' $M_0$, 
the system exhibits a nonequilibrium scaling regime in which the order parameter typically 
shows an anomalous power-law increase as \cite{janssen} 
\be
M(t) \sim M_0 \, t^{\theta}
\ee
before eventually relaxing toward its equilibrium value, where the exponent $\theta$ is 
a new nonequilibrium critical exponent. Though it typically takes a positive value, it 
becomes negative for some systems such as the tricritical system \cite{janssen2}, 
the Ashkin-Teller model \cite{llsz}, the Blume-Capel model \cite{das3}, the $4$-state Potts 
model \cite{das2}, and for the BW model \cite{santos,arashiro,malakis}. The time scale 
$\tau_M$ associated with the power-law increase depends upon the magnitude of the initial 
magnetization as $\tau_M \sim M_0^{-z/x_0}$ where the scaling dimension $x_0$ is related 
to the other exponents as $x_0=\theta z + \beta \nu$.

In practice, $\theta$ was obtained by linearly extrapolating the effective exponents 
$\theta_{eff}$ (for finite $M_0$) vs. $M_0$ to $M_0=0$. Alternative way of obtaining 
$\theta$ was first proposed by Huse \cite{huse}. Using a scaling ansatz for the nonequilibrium 
structure factor of the two-dimensional kinetic Ising model, Huse has shown that the 
auto-correlation function of the global order-parameter ${\cal M}(t)$ with random initial 
conditions exhibits a power-law behavior 
\be
{\cal A}_{\cal M}(t) =\frac{1}{N} < {\cal M}(t) {\cal M}(0) > \sim t^{\theta}
\ee
Huse has also shown that the spin auto-correlation function $A(t)$ exhibits 
a critical decay in time as 
\be
A(t) \equiv \frac{q}{q-1} \frac{1}{N}
<\sum_i \Big( \delta(\sigma_i(0), \sigma_i(t))-\frac{1}{q} \Big) > 
\sim t^{-\lambda} \equiv t^{-\frac{d}{z} + \theta}
\ee
Tom\'e and Oliveira \cite{tome1} have proved that (18) holds for the equilibrium as well 
as nonequilibrium systems (i.e. without detailed balance) possessing the global up-down 
symmetry. Later on, Tom\'e \cite{tome2} has shown that the validity of (18) is extended to
systems with other symmetries. It is much more convenient to measure $\theta$ by measuring 
the auto-correlation ${\cal A}_{\cal M}(t)$ since one is free from cumbersome preparation 
of the initial states with very small magnetization and an extrapolation procedure to 
vanishing initial magnetization. 

The new nonequilibrium exponent $\theta$ has been obtained as follows. For $q=2$,  
$\theta=0.191(1)$ (Heat Bath) and $\theta=0.197(1)$ (Metropolis) using (16) 
\cite{okano,bz1,zhang,ying}, and $\theta=0.19$ using (17) \cite{huse}. In addition, 
Grassberger \cite{grassberger} also measured the exponent $\theta$ as $\theta=0.191(3)$ 
using a damage spreading method. Likewise, for $q=3$,  The exponent $\theta$ takes the 
following values: $\theta=0.075(3)$ (Heat Bath) and $\theta=0.070(2)$ (Metropolis) using 
(16) \cite{bz1,okano,zhang,ying}, and $\theta=0.072(1)$ using (17) \cite{das2}. 
The exponent $\theta$ was also measured using (17) as $\theta=0.093(4)$ \cite{brunstein} 
for a nonequilibrium cellular automaton model with $C_{3v}$ symmetry which belongs to 
the $3$-state Potts universality class \cite{tome3}. In addition, there exists measured 
value of $\theta$ for $q=4$ using both (16) and (17), which are given respectively by 
$\theta=-0.0471(33)$ and $\theta=-0.0429(11)$ \cite{das2}. It is an interesting fact that  
the $4$-state Potts model and the BW model, which belong to the same static universality 
class, do not share the same nonequilibrium exponent $\theta$: 
$\theta=-0.185(2)$ \cite{malakis}  or $-0.186(2)$ \cite{arashiro} for the BW model. 
It would be interesting as well to clarify the question as to whether this nonuniversal 
dynamic behavior is extended to the case of $z$ in both systems. 
 
Independent measurements of the exponents $\theta$ and $\lambda$ using (17) and (18) provide 
an alternative way \cite{bz1,schzh1,okano} of obtaining $z$ through the relation 
 $z=d/(\lambda+\theta)$. The measured values of $\lambda$ and hence $z$ are given by  
$\lambda=0.737(1)$ and $z=2.155(3)$ for $q=2$, and $\lambda=0.836(2)$ and $z=2.196(8)$ for 
$q=3$ \cite{bz1}. As for the BW model, $\lambda=1.188(10)$ and $z=1.994(24)$ \cite{malakis} 
were obtained using the continuous-time Monte Carlo algorithm. One disadvantage of this method 
is that the rather large value of $\lambda$ induces strong fluctuations in its measurement, 
which is the main source of the error for the estimate of $z$. 

With $z$ in hand, one can use (11) with disordered initial states to obtain the ratio 
$\beta/\nu$, or equivalently, the exponent $\eta \equiv 2 \beta/\nu$ (for $d=2$). In order 
to determine $\beta$ and $\nu$ separately, one can use the following scaling ansatz for the 
order parameter near and below $T_c$ 
\be
M(t, \epsilon)=t^{-\beta/z\nu} {\cal F}(t^{1/z\nu} \epsilon)
\ee
where ${\cal F}$ is the scaling function, and $\epsilon$ is the reduced temperature defined as 
$\epsilon \equiv (T_c-T)/T_c$. The equation (19) reduces to (4) for $\epsilon=0$. Then one can 
give an  estimate for $1/\nu$ using
\be
\frac{\partial}{\partial \epsilon} \ln M(t, \epsilon)|_{\epsilon
=0}  = t^{1/z\nu} \Big( \frac{d}{dx} \ln {\cal F}(x)\Big)_{\epsilon=0}
\ee 
This method gives $1/\nu=1.03(2)$ (square lattice) \cite{bz1} and $1/\nu=1.027(6)$ (triangular 
lattice) \cite{ying} for $q=2$, and $1/\nu=1.24(3)$ (square lattice) \cite{bz1,schzh2}
and $1/\nu=1.223(8)$ (triangular lattice) \cite{ying} for $q=3$. But, involving the difference 
of the order parameter at several close temperatures near $T_c$ may cause considerable error in 
the estimate of the exponent $\nu$. Our method (pesented in Section III.C) utilizing the energy 
relaxation can provide an alternative and better way of estimating the exponent $\nu$ without 
using (20).

\section{Summary and concluding remarks}
 
Through Monte Carlo simulations we have investigated the nonequilibrium critical dynamics of 
the ferromagnetic $q$-state Potts model on a square lattice. Primary purpose of the present 
work was to evaluate how accurate estimates the NER method can provide for both static and 
dynamic exponents of the statistical systems. The ferromagnetic Potts model is an ideal test 
system for that purpose since the static critical properties of the model are exactly known. 

In contrast to other nonequilibrium approaches, we  utilized both the order parameter and
energy relaxation. The ratio of the power-law exponents of the order parameter and the energy 
involves solely the static exponents. In order to separately measure the two independent 
static exponents, one first measures the dynamic exponent $z$ by considering the 
order-parameter moments starting from either random or fully ordered initial states. It is 
found that in the present model, when the fully ordered initial states are employed, the time 
scale associated with the asymptotic scaling for the relaxation of the second moments of 
the order parameter sets in, seems to be much longer for higher $q$ than that for the disordered 
initial states and may exceed the present simulation time window. As a result, the slopes
obtained from the second moments of the order parameter may differ. One thus has to 
be very careful in obtaining the dynamic exponent $z$ from the time-dependent moments 
for the order parameter. Once $z$ is determined, the two independent exponents $\nu$  and  
$\beta$ can be determined from the relaxation of the energy and the order parameter, 
respectively. 

The present work has demonstrated that, in the case of $q=2$ (Ising) model this method can 
provide accurate estimates for the static and dynamic exponents provided that, as mentioned 
above, special care is taken in the choice of the initial states for the estimation 
of the dynamic exponent $z$. For example the dynamic exponent $z_{mixed}$ obtained from 
the relaxation of the second moment of the order parameter with disordered initial states 
are more consistent with the relaxation of the order parameter and excess energy when the 
exact values of the static exponents are assumed. 
In the case of $q=3$ Potts model, even though the static exponents obtained with the same 
methods are less accurate than in the case of $q=2$, they still show reasonable agreement
with the exact known values of the static exponents.  

The fact that the dynamic exponent $z$ does not appear in the ratio of the power-law exponents 
of the order parameter and energy, may be used to clarify the nature of the phase transition 
whose critical properties are under debate. An outstanding example is the phase transition 
associated with the chirality order in the FFXY models. The relaxations of the order parameter
and energy are expected to become slower for frustrated systems. Hence the above ratio can be 
measured  with higher accuracy. We therefore tend to believe that once the critical temperature 
is accurately determined, the critical relaxations of the order parameter and energy for the 
chirality phase transition can tell whether the phase transition belongs to the Ising 
universality class, or to the $3$-state Potts class, or else to an entirely different 
universality class. 

\acknowledgements

We are grateful to Prof. Jooyoung Lee for generously allowing us to access the Gene Cluster at
at Korea Institute for Advanced Study (KIAS) where part of the computations were carried out.

\appendix*

\section{Critical relaxations of moments of the order parameter and energy}
The critical relaxations of the order parameter and energy, and their higher moments can be 
obtained from the following scaling behavior of the nonequilibrium generating function 
$\psi(\epsilon, h, t)$ which reduces to the equilibrium free energy density in the limit 
$t\rightarrow \infty$:
\bea
\psi(\epsilon, h, t)&=& b^{-d} {\tilde \psi}(\epsilon \cdot b^{y_T},
 h \cdot b^{y_h}, t \cdot b^{-z}) 
\nonumber \\
&=& t^{-d/z} {\tilde \psi}(\epsilon \cdot t^{y_T/z}, h \cdot t^{y_h/z}, 1)
\eea
where $b$ is the scaling factor, $\epsilon \equiv |T_c - T|/T_c$ the reduced temperature 
difference, and $h$ the external (magnetic) field coupled to the order parameter.
The above result was first derived by Suzuki \cite{suzuki}. In (A.1), $y_T$ and $y_h$ are 
the scaling dimensions associated respectively with temperature and external field, which 
are given by $y_T=1/\nu$ and $y_h=d-\beta/\nu$. The relaxation of the order parameter 
$M(t)$ at $T_c$ is obtained by differentiating the generating function $\psi(\epsilon, h, t)$
with respect to the magnetic field $h$:
\be
M(t)=\Big( \frac{\partial \psi}{\partial h}\Big)_{\epsilon=h=0} \sim t^{(y_h-d)/z}
=t^{-\beta/z \nu}
\ee
Likewise, the energy relaxation is obtained from the derivative of the generating function 
with respect to the reduced temperature
$\epsilon$:
\be
\big(E(t)-E_c \big) = \Big( \frac{\partial \psi}{\partial \epsilon}\Big)_{\epsilon=h=0} 
 \sim  t^{(y_T-d)/z}=t^{-(\nu d-1)/\nu z}=t^{-(1-\alpha)/z \nu}
\ee
where the hyperscaling relation $2-\alpha=\nu d$ was used in the last equality.
The equations (A.2) and (A.3) gives the leading order part of (4).

The second order moment of the order-parameter is accordingly obtained by differentiating
twice $\psi (\epsilon, h, t)$ with respect to $h$:
\be
< (\delta {\cal M}(t))^2 > =<{\cal M}^2(t)>-M^2(t)=
\Big( \frac{\partial^2 \psi}{\partial h^2}\Big)_{\epsilon=h=0}
\sim t^{(2 y_h -d)/z}=t^{(d-2\beta/\nu)/z}
\ee
The equation (A.4) gives (11).

Likewise, the second-order moment of the energy is given by
\be
<(\delta {\cal E}(t))^2> =
\Big( \frac{\partial^2 \psi}{\partial \epsilon^2}\Big)_{\epsilon=h=0}
\sim t^{(2y_T-d)/z}=t^{(2-\nu d)/z \nu}=t^{\alpha/z \nu} 
\ee
The equation (A.5) gives the last member of (15).

Finally, the cross-moment of the order parameter and energy is given by
\be
<\delta {\cal M}(t) \delta {\cal E}(t)> =
\Big( \frac{\partial^2 \psi}{\partial \epsilon 
\partial h}\Big)_{\epsilon=h=0} \sim 
t^{(y_T+y_h-d)/z} = t^{(1-\beta)/z\nu} 
\ee
Dividing $<\delta {\cal M}(t) \delta {\cal E}(t)>$ by $M(t)$ gives
\be
\frac{<\delta {\cal M}(t) \delta {\cal E}(t)>}{M(t)} \sim t^{1/z \nu}
\ee
The equation (A.7) yields the first member of (15).






\bigskip

\newpage 
\centerline {\bf FIGURE CAPTIONS}
 
\renewcommand{\theenumi}{Fig.~1}
\begin{enumerate}
\item
The relaxations of the order parameter $M(t)$ for (a) $q=2$ 
and for (b) $q=3$ at temperatures near $T_c$ 
((a) $T=1.130$, $1.1323$, $1.1342$, $1.13459 (T_c)$, $1.135$, $1.137$, $1.14$, and 
(b) $T=0.993$, $0.994$, $0.9945$, $0.99497 (T_c)$, $0.9955$, $0.996$, $0.997$).
The critical temperature can be  determined as the temperature at which $M(t)$ 
exhibits a critical relaxation (represented by  a broken line) through a change 
from an upward curvature to a downward curvature.
\end{enumerate}

\renewcommand{\theenumi}{Fig.~2}
\begin{enumerate}
\item
The relaxations of the energy difference $\big(E^{\ast}-E(t) \big)$ 
in a log-log plot for (a) $q=2$ and (b) $q=3$ with various test values of $E^{\ast}$
((a) $E^{\ast}=-1.704$, $-1.705$, $-1.706$, $-1.70711 (E_c)$, $-1.708$, $-1.709$, $-1.710$, 
and (b) $E^{\ast}=-1.571$, $-1.573$, $-1.575$, $-1.57735 (E_c)$, $-1.579$, $-1.581$, 
$-1.583$).  
By tuning $E^{\ast}$ such that $\big(E^{\ast} - E(t) \big)$ exhibits the best power-law 
relaxation, one can determine both $E_c$ and $-(\nu d -1)/z\nu$.  For each figure, 
the broken line represents the relaxation with the exact critical energy $E_c$.
\end{enumerate} 

\renewcommand{\theenumi}{Fig.~3}
\begin{enumerate}
\item
(a) The critical relaxation of the order parameter $M(t)$, and (b) the  critical relaxation 
of the energy difference $\big(E_c - E(t) \big)$.
In (a) and (b) dot-dashed lines are curves obtained from FIT3A with optimal fits. 
Shown in (c) and (d) are plots of $M_A(t)\equiv M(t) A_0 t^{\beta/z\nu}$ (filled square) 
together with the fitting function (solid line) for $q=2$ and $q=3$ respectively.
Here $A^{-1}_0 t^{-\beta/z \nu}$ represents the asymptotic scaling obtained from the fitting.  
Shown in (e) and (f) are the plots of $E_A(t)\equiv \big(E_c - E(t) \big) B_0 t^{(\nu d -1)/z\nu}$ 
vs. $t$ (filled square) together with the fitting function (solid line) for $q=2$ and $q=3$ 
respectively. Here $B^{-1}_0 t^{-(\nu/d -1)/z \nu}$ represents the asymptotic scaling  
obtained from the fitting.  

\end{enumerate} 

\renewcommand{\theenumi}{Fig.~4}
\begin{enumerate}
\item
The order-parameter cumulant $f_{{\cal M}{\cal M}}(t)$ vs. $t$ in a log-log 
plot for $q=2$ (a) and $q=3$ (b). 
(c): $ f_{{\cal M}{\cal M}, A}(t)\equiv f_{{\cal M}{\cal M}}(t) C_{0} t^{-d/z}$ vs. $t$ 
in the case of mixed initial states for $q=2$ (filled square) together with the fitting
function (solid line). Here $C^{-1}_0 t^{d /z}$ represents the asymptotic scaling 
obtained from the fitting.  
Dotted lines are the straight lines indicating the asymptotic power law behavior.
\end{enumerate}

\renewcommand{\theenumi}{Fig.~5}
\begin{enumerate}
\item
The second-order moments $f_{{\cal M}{\cal E}}(t)$ (a) and $f_{{\cal E}{\cal E}}(t)$ (b) 
vs. $t$ in a log-log plot for $q=2$ and $q=3$. (c) $f_{{\cal E}{\cal E}}(t)$ vs. $\ln t$ 
in  a log-log plot for $q=2$. This plot shows a logarithmic time dependence of 
$f_{{\cal E}{\cal E}}(t)\sim (\log t)^{\phi}$ with $\phi \simeq 0.80$ for the long-time 
region ($t \ge 100$ (mcs)).
\end{enumerate}

\newpage 

\begin{table}[h]\centering
\begin{tabular}{|c|c|ccccc|} \hline \hline
        & &    $z$      &         $1/\nu$        & $\beta$  
                    &  $\eta = 2 \beta/\nu$ ($d=2$) & $\alpha=2-\nu d$ \\
\cline{2-7}
 $q=2$  & \mbox{present work} &  2.1668(19)($z_{MM}$)    &   0.9984(25)  & 0.12527(51)  
                 & 0.2501(16)   & -0.0032(50)       \\
\cline{2-7}
  & \mbox{short-time dynamics} & 2.155(3) & 1.03(2) & 0.1165 &  0.240(15)&0.058(38) \\
 \cline{2-7}
 & \mbox{exact}&  &    1    &  1/8=0.125   &  1/4=0.25    &0  \\ 
\hline \hline

  &\mbox{present work} &  2.1735(40) ($z_{MM,disordered}$)   & 1.213(6)   & 0.1080(20)
                 &  0.2618(83) & 0.350(22)\\
\cline{2-7}
$q=3$ & \mbox{short-time dynamics} & 2.196(8) & 1.24(3) & 0.1085 &  0.269(7)&0.387(40)\\
\cline{2-7}
& \mbox{exact}&  & 6/5=1.2 & 1/9=0.1111 & 4/15=0.2666 & 1/3=0.3333\\
\hline\hline
\end{tabular}
\caption{The measured dynamic and static critical exponents.}
\label{t1}
\end{table}

\begin{table}[h]\centering
\begin{tabular}{|c|c|c|c|c|} \hline \hline
      & $c''_M$  & $ \beta /\nu z $  &  $\delta_M$ & $z_{M}$       \\
\cline{2-5}
    &  0.0   &  0.0576815    &  0.8956     & 2.16707       \\
\cline{2-5}
    &  0.01   & 0.0576911   & 0.9165      & 2.16671      \\
 \cline{2-5}
    &  0.02    & 0.0577003   & 0.9366     & 2.16637     \\ 
 \cline{2-5}
\mbox{FIT3A} &  0.03  & 0.0577094 & 0.9559  & 2.16603    \\ 
 \cline{2-5}
      &  0.04    & 0.0577181   & 0.9745    &  2.16570   \\ 
 \cline{2-5}
      &  0.05    & 0.0577266   & 0.9924   & 2.16538     \\ 
 \cline{2-5}
      &  0.07    & 0.0577427   & 1.0265  & 2.16478     \\ 
 \cline{2-5}
      &  0.1    & 0.0577652   & 1.0738    & 2.16393  \\ 

\hline \hline

         &  0.0  & 0.05768   & 0.907    & 2.1671         \\
\cline{2-5}
          & 0.01 & 0.057689  & 0.9264  & 2.16679          \\
\cline{2-5}
          & 0.02 & 0.057699  & 0.9442   & 2.16642          \\
\cline{2-5}
          & 0.03 & 0.057708  & 0.9612  & 2.16606          \\
\cline{2-5}
\mbox{FIT3C} & 0.04 & 0.057717 & 0.9774 & 2.16573         \\
\cline{2-5}
          & 0.06  & 0.0577338  & 1.0078  & 2.16511          \\
\cline{2-5}
          & 0.08  & 0.0577494 & 1.0363  & 2.16453          \\
\cline{2-5}
          & 0.1   & 0.0577642 & 1.0633 &  2.16397          \\

\cline{2-5}
\mbox{FIT3B} & 0.07834 & 0.05769  & 1.0 (fixed)  & 2.16675          \\
\hline\hline
\end{tabular}
\caption{The dominant exponents and the first correction exponents for the relaxation of
magnetization, obtained from the two fitting procedures FIT3A and FIT3C for the case 
of $q=2$. $z_M $ denotes the value of the dynamic exponent derived from the value of the 
second column for $\beta / z \nu$ using the exact values of the static exponents for
$\beta $ and $\nu$. $c''_M$ represents the chosen values of the coefficients of the highest 
order correction terms for fits. For comparison, we also added (the last line) the
result of a fit with FIT3B where the first correction exponent $\delta_M$ is set equal to 
unity }
\label{t2}
\end{table}

\begin{table}[h]\centering
\begin{tabular}{|c|c|c|c|c|} \hline \hline
      &   $\delta_E $ & $(\nu d -1) /\nu z$  & $c''_E$ & $z_E$  \\
\cline{2-5}
             &  0.90   &  0.46201   &   0.2028    & 2.1645      \\
\cline{2-5}
 \mbox{FIT3A}  &  0.95   &  0.46227   &  0.47124  & 2.1632    \\
 \cline{2-5}
             &  1.0    & 0.46253   & 0.76616   & 2.1620    \\ 

\hline \hline

           & 0.93  & 0.462169  & 0.26146   & 2.1637         \\
\cline{2-5}
           & 0.95  & 0.462265  & 0.40190  & 2.1633          \\
\cline{2-5}
\mbox{FIT3C} & 0.96  & 0.462315 & 0.47326  & 2.1630         \\
\cline{2-5}
           & 0.97 & 0.462367    & 0.54537  & 2.1628          \\
\cline{2-5}
           & 1.0 & 0.462530    & 0.766162  & 2.1620         \\
\hline\hline
\end{tabular}
\caption{The dominant exponents and the first correction exponents for the energy
relaxation, obtained from the two fitting procedures FIT3A and FIT3C for the case of 
$q=2$. Note that $z_E $ denotes the value of the dynamic exponent derived from the 
value of the second column for $(\nu d -1)/\nu z$ using the exact values of the 
static exponents for $\nu$. Here the exponents  $\delta_E$'s represent the chosen 
values of the first correction exponents for fitting. }
\label{t3}
\end{table}

\begin{table}[h]\centering
\begin{tabular}{|c|c|c|c|c|} \hline \hline
      & $c''_M$  & $ \beta /\nu z $  &  $\delta_M$ & $z_{M}$  \\
\cline{2-5}
   &  0.03   &  0.0600345     &  0.6415      & 2.2209      \\
\cline{2-5}
       &  0.06   & 0.0601009   & 0.7063      & 2.2185     \\
 \cline{2-5}
      &  0.07    & 0.0601192   & 0.7260      & 2.2178     \\ 
 \cline{2-5}
\mbox{FIT3A} &  0.08  & 0.0601374  & 0.7451 & 2.2171     \\ 
 \cline{2-5}
      &  0.09    & 0.060154   & 0.7634     & 2.2165     \\ 
 \cline{2-5}
      &  0.10    & 0.060170   & 0.7811     & 2.2159     \\ 
 \cline{2-5}
      &  0.12    & 0.0601995   & 0.8149    & 2.2149  \\ 
 \cline{2-5}
      &  0.15    & 0.060241   & 0.8630     & 2.2133   \\ 

\hline \hline

         &  0.06  & 0.060141  & 0.7713  & 2.2170     \\
\cline{2-5}
          & 0.09 & 0.060183  & 0.8139  & 2.2155     \\
\cline{2-5}
          & 0.10 & 0.0601959 & 0.8274  & 2.2150      \\
\cline{2-5}
\mbox{FIT3C} & 0.12 & 0.0602175 & 0.8524  & 2.2142     \\

\cline{2-5}
          & 0.15 & 0.060249 & 0.88845 & 2.2130 \\ 
\cline{2-5}
FIT3B    & 0.244  & 0.0603559 & 1.0 (fixed)  & 2.2091 \\
\cline{2-5}
FIT0     & -      & 0.06058   & -     & 2.2010        \\
\hline\hline
\end{tabular}
\caption{The dominant exponents and the first correction exponents for the relaxation of
magnetization, obtained from the two fitting procedures FIT3A and FIT3C for the case 
of $q=3$. $c''_M$ represents the chosen values of the coefficients of the highest order 
correction terms for fits. Also shown in the last two lines are the results of
fits using FIT3B and FIT0 (with no correction terms) respectively. }
\label{t4}
\end{table}

\begin{table}[h]\centering
\begin{tabular}{|c|c|c|c|c|} \hline \hline
      &   $\delta_E $ & $(\nu d -1) /\nu z$  & $c''_E$ & $z_E$ \\
\cline{2-5}
             &  0.55  &  0.36008  &   0.3898 & 2.2218  \\ 
\cline{2-5}
 \mbox{FIT3A}  &  0.60  & 0.36142 &  0.6459  & 2.2135  \\  
 \cline{2-5}
             &  0.61    & 0.36167 &  0.6997  & 2.2119  \\ 
 \cline{2-5}
             &  0.62    & 0.36192 &  0.7544  & 2.21045 \\ 
 \cline{2-5}
             &  0.63    & 0.36216 &  0.8099  & 2.20899  \\ 
 \cline{2-5}
             &  0.64    & 0.36239 &  0.8664  & 2.20756  \\ 

\hline \hline

            & 0.70  & 0.36298  & 0.7896  & 2.20397      \\
\cline{2-5}
\mbox{FIT3C} & 0.72  & 0.36336  & 0.9676  & 2.20165     \\

\cline{2-5}
             & 0.73    & 0.36356  & 1.057  & 2.2005     \\
\cline{2-5}
            & 0.74    & 0.36375  & 1.1464 & 2.1993 \\

\hline\hline
\end{tabular}
\caption{The dominant exponents and the first correction exponents for the energy
relaxation, obtained from the two fitting procedures FIT3A and FIT3C for the case of 
$q=3$. Here the exponents  $\delta_E$ represent the chosen values of the first
correction exponents for fits. }
\label{t5}
\end{table}

\begin{table}[h]\centering
\begin{tabular}{|c|c|c|c|c|} \hline \hline
      & $\delta_{MM} $ & $c''_{MM}$  & $d/z$ & $z_{MM}$ \\
\cline{2-5}
            & 1.35  &    0.0   & 0.92323 & 2.16631   \\
\cline{2-5}
            &  1.3  &    0.0  &  0.92311 & 2.16659  \\
 \cline{2-5}
            &  1.25  &   0.0  & 0.92298 & 2.16689    \\ 
 \cline{2-5}
\mbox{FIT3A} &  1.20  &   0.0 & 0.92285 & 2.16720    \\ 
 \cline{2-5}
            &  1.15  &   0.0 & 0.92271 & 2.16753     \\ 
 \cline{2-5}
            &  1.0   & 0.6923 & 0.92314 & 2.16652   \\ 
 \cline{2-5}
            &  1.0   & 0.0 & 0.92224 & 2.16863    \\ 

\hline \hline
\mbox{FIT3C} & 1.34595 & 0.0  & 0.92335  & 2.16603  \\

\hline\hline
\end{tabular}
\caption{The dominant exponents and the first correction exponents for the 
relaxation of the second moment of magetization with {\em random} initial states, 
obtained from the two fitting procedures FIT3A and FIT3C for the case of $q=2$. 
Here, the fitting is mostly done up to second corrections. $c''_M$ is the 
chosen values of the coefficients of the highest order correction terms for 
fits.}
\label{t6}
\end{table}

\begin{table}[h]\centering
\begin{tabular}{|c|c|c|c|c|} \hline \hline
      & $\delta_{MM} $ & $c'_{MM}$  & $d/z$ & $z_{MM}$ \\
\cline{2-5}
           & 0.8317 &  0.15  & 0.927756  & 2.15574      \\
\cline{2-5}
           & 0.8471 & 0.09  & 0.927876 & 2.15546        \\
 \cline{2-5}
            & 0.8545 & 0.06 & 0.927933 & 2.15533         \\ 
 \cline{2-5}
\mbox{FIT2A}& 0.8618 & 0.03 & 0.927988 & 2.15520          \\ 
 \cline{2-5}
            & 0.8689 & 0.0 & 0.928042 & 2.15507       \\ 
 \cline{2-5}
           & 0.8713 & -0.01 & 0.92806 & 2.15503        \\ 

\hline \hline
\mbox{FIT0} & - & -  & 0.92819  & 2.15473  \\  

\hline\hline
\end{tabular}
\caption{The dominant exponents and the first correction exponents for the 
relaxation of the second moment of magetization with {\em ordered} initial states, 
obtained from the two fitting procedures FIT2A and FIT0 for the case 
of $q=2$. The column for $c'_M$ lists the chosen values of the coefficients of 
the second correction terms for fits.}
\label{t7}
\end{table}


\begin{thebibliography}{99}
\bibitem{ffxy}
It is well known that the unfrustated XY model in two-dimension 
undergoes a Berezinskii-Kosterlitz-Thouless (BKT) 
transition \cite {berez,kt} at a finite temperature due to the unbinding 
of bound vortex pairs. 
As in the unfrustrated XY model, the FFXY model also exhibits 
a BKT-like transition at a finite temperature. 
The FFXY model possesses additional discrete Ising-like symmetry 
corresponding to the double degeneracy of the chirality 
configurations as well as continuous rotational symmetry, 
leading to the new Ising-like phase transition at finite temperature. 
First controversy is  whether the two transitions take place at
exactly the same temperature, or at close but distinct temperatures.
Numerically, the two transition temperatures are found to be very close to each other.
Related to this is another controversy on the nature of the two
phase transitions in the FFXY models: whether or not the BKT-like
 transition has the same critical properties as 
the genuine BKT transition of the unfrustrated XY
model, and whether or not the chirality transition has the same 
critical properties as the ordinary two-dimensional Ising model.
\end{thebibliography}
\end{document}